\documentclass[twocolumn,showpacs,preprintnumbers,amsmath,amssymb]{revtex4}
\usepackage{epsfig,subfigure}
\usepackage{graphicx}
\usepackage{amsmath}
\usepackage{color}

\newcommand{\etal}{\emph{et al.}}

\begin{document}

\preprint{FIS-UI-TH-04-02}

\def\Journal#1#2#3#4{{#1} \textbf{#2}, #3 (#4)}
\def\RPP{{Rep. Prog. Phys.}}
\def\PRC{{Phys. Rev. C}}
\def\PRD{{Phys. Rev. D}}
\def\FP{{Found. of Physics}}
\def\ZPA{{Z. Phys. A}}
\def\NPA{{Nucl. Phys. A}}
\def\JPG{{J. Phys. G: Nucl. Part}}
\def\PRL{{Phys. Rev. Lett.}}
\def\PRpt{{Phys. Report.}}
\def\PLB{{Phys. Lett. B}}
\def\AP{{Ann. Phys. (N.Y.)}}
\def\EPJA{{Eur. Phys. J. A}}
\def\NP{{Nucl. Phys.}}
\def\PR{{Phys. Rev.}}
\def\RMP{{Rev. Mod. Phys.}}
\def\IJMPE{{Int. J. Mod. Phys. E}}
\input epsf

\title{ Neutron Fraction and Neutrino Mean Free Path Predictions in Relativistic Mean Field Models }

\author{P.T.P. Hutauruk, C.K. Williams, A. Sulaksono, T. Mart}
\affiliation{Departemen Fisika, FMIPA, Universitas Indonesia, Depok
16424, Indonesia}

\begin{abstract}
 The equation of state (EOS) of dense matter and neutrino mean free path (NMFP) in a neutron star have been studied by using relativistic mean field models motivated by effective field theory (ERMF). It is found that the models predict too large proton fractions, although one of the models (G2) predicts an acceptable EOS. This  is caused by the isovector terms. Except G2, the other two models predict anomalous NMFP. In order to minimize the anomaly, besides an acceptable EOS, a large $M^*$ is favorable. A model with large $M^*$ retains the regularity in the NMFP even for a small neutron fraction.
\end{abstract}


\maketitle
The finite-range (FR) (see Refs.~\cite{pg,ring,serot,sil}) and point-coupling (PC)(see Refs.~\cite{niko,rusnak,buerven,anto1,anto2}) types of relativistic mean field (RMF) models have been quite successful to describe the bulk as well as single particle properties in a wide mass spectrum of nuclei.

The early version of RMF-FR is based on a Lagrangian density which uses nucleon, sigma, omega and rho mesons as the degrees of freedom with additional cubic and quartic nonlinearities of sigma meson. For example NLZ, NL1, NL3 and NL-SH parameter sets belong to this version. Recently, inspired by the effective field and density functional theories for hadrons, a new version of this model (ERMF-FR) has been constructed~\cite{serot,furnstahl}. It has the same terms like the previous RMF-FR but with additional isoscalar and isovector tensor terms and nonlinear terms in the form of sigma, omega and rho mesons combination. One of parameter sets of this version is G2. Besides yielding accurate predictions in finite nuclei and normal nuclear matter~\cite{serot,furnstahl,sil}, G2 has the demanding features like a positive value of quartic sigma meson coupling constant that leads to the existence of lower bound in energy spectrum of this model~\cite{baym,arumu} and to the missing zero sound mode in the high density symmetric nuclear matter~\cite{cailon}. Moreover, the agreement of the nuclear matter and the neutron matter equation of states (EOS) in high density of G2 with the Dirac Brueckner Hartree Fock (DBHF) calculation ~\cite{sil,arumu} is better than those of NL1, NL3 and TM1 models (the standard RMF-FR plus a quartic omega meson interaction). 

The difference between RMF-PC and RMF-FR is due to the replacement of mesonic interactions in the FR model by density dependent interactions. It is evident that RMF-PC and RMF-FR serve similar quality in predicting finite nuclei and normal nuclear matter~\cite{niko,buerven,anto2}. This is due to the fact that ``finite-range'' effects in RMF-PC model are effectively absorbed by the coupling constants. Therefore in connection with different treatments of ``finite-range'' in both models, studying the behavior of the PC model in high density should be interesting. In this report, we choose the VA4 parameter set of Ref.~\cite{rusnak} (ERMF-PC)  because it can be properly extrapolated to the high density and it has also density dependent self- and cross-interactions in the nonlinear terms.

So far the EOS of a neutron star has not been known for sure~\cite{reddy1}. However, recently~\cite{daniel} the flow of matter in heavy ion collisions has been used to determine the pressure of nuclear matter with a density from 2 until 5 times the nuclear saturation density ($\rho_0$). Reference~\cite{daniel} has found that these data can be explained only by the variational calculation of Akmal \etal~\cite{akmal}. Unfortunately, this interaction cannot be successfully applied to the case of finite nuclei~\cite{arumu}. Reference~\cite{arumu} found that the EOS predicted by G2 is in agreement with data. This result is remarkable, since Ref.~\cite{horo1} states that the minimal requirement for an accurate neutrino mean free path (NMFP) is a correct prediction in the low density limit, as well as the consistency with the corresponding EOS. On the other hand, one should remember that  many-body corrections are important but they depend on the model and the approximation of strong interaction used~\cite{horo1,mornas1,mornas2,reddy1,reddy2,horowitz91,niembro01,yama,caiwan,margue}.

According to Refs.~\cite{blaschke,kolo} all RMF-FR models yield lower threshold densities for direct URCA process than those of variational calculations~\cite{akmal}. In the neutron star cooling model, Migdal \etal ~\cite{migdal} treated this fact as a fragile point of RMF-FR models. So, they disregarded direct URCA from their analysis but Lattimer \etal~\cite{lati} used this fact to develop their direct URCA scenario.

Therefore, in this report we will compare the neutron matter prediction in high density from the G2, NLZ and VA4 models in order to check the result of Ref.~\cite{arumu} and  the possibly different predictions from ERMF-PC and ERMF-FR due to the different treatment of the ``finite-range effects''. Furthermore, the agreement between the G2 EOS with experimental data has motivated us to calculate NMFP using this model for direct URCA process. A similar assumption as in Ref.~\cite{niembro01} is used, i.e., the ground state of the neutron star is reached once the temperature has fallen below a few MeV. This state is gradually reached from the later stages of the cooling phase. The system is then quite dense and cool so that zero temperature is valid. In this case the direct URCA neutrino-neutron scattering is kinematically possible for low energy neutrinos at and above the threshold density when the proton fraction exceeds 1/9~\cite{lati} or slightly larger if muons are present. Furthermore, the absorption reaction is suppressed. For simplicity, we neglect the RPA correlations.

\begin{figure}[b]
\centering
\begin{tabular}{c@{\qquad}c}
 \mbox{\epsfig{file=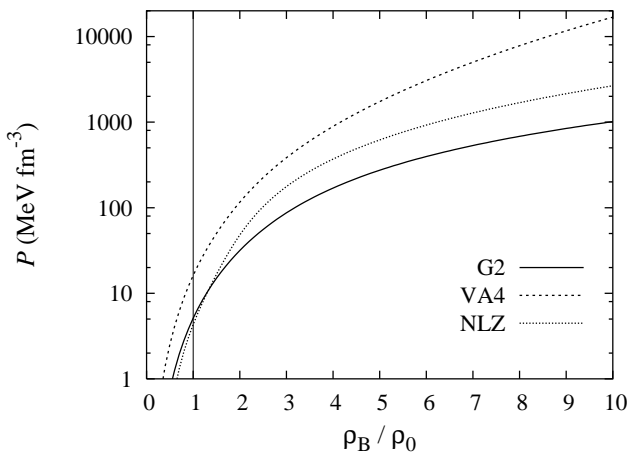,height=6.2cm}}
\end{tabular}
\caption{Equation of states (EOS) of the neutron matter.}\label{fig1}
\end{figure}

\begin{figure}[t]
\centering
\begin{tabular}{c@{\qquad}c}
 \mbox{\epsfig{file=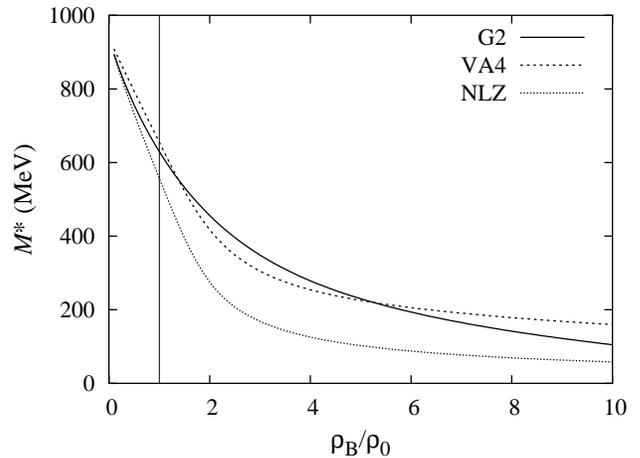,height=6.2cm}}
\end{tabular}
\caption{Effective masses ($M^*$) of the neutron matter.}\label{fig1b}
\end{figure}

The effects of self- and cross-interactions terms  and the treatment of finite-range in high density can be  observed by extrapolating the EOS which is presented by the neutron matter pressure $P$ and the effective mass $M^*$, as shown in Figs.~\ref{fig1} and~\ref{fig1b}, where we compare the results obtained from the G2~\cite{furnstahl}, NLZ \cite{pg}, and VA4~\cite{rusnak}  models as a function of $\rho_B/\rho_0$.

It is found that the nuclear matter EOS of VA4 is stiffer than those of NLZ and G2, even for $\rho_B$ less than $\rho_0$.  However, the G2 EOS is softer than the NLZ one at the high density but not at the low density. This fact emphasizes the result of Ref.~\cite{arumu} that  the crucial role of self- and cross-interactions of meson exchange model is to soften the EOS at the high density. 

 It is shown in Fig.~\ref{fig1b} that for 1 $\leq$ ${\rho_B}/{\rho_0}$ $\leq$ 5, the effective mass $M^*_{G2}$ $>$  $M^*_{VA4}$, but for ${\rho_B}/{\rho_0}$ $\geq$ 5 one observes that $M^*_{G2}$ $<$  $M^*_{VA4}$. This indicates that quantitatively $M^*$ depends on the model. We note here that the effective masses of G2 and VA4 depend on self- and cross-interaction terms implicitly. We also note that other mechanisms could also produce a larger $M^*$, e.g.,  in the Zimanyi-Moszkowski  and linear Hartree-Fock Walecka models~\cite{niembro01}, where those terms are not present. Although those models give  a regular NMFP, they  are quite unsuccessful in finite nuclei applications, especially in predicting the single particle spectra of nuclei~\cite{hua}. Therefore, it is interesting to check whether or not the relation between a large $M^*$ and a regular NMFP also appears in the case of ERMF models.

Now, we calculate the NMFP of the neutron star matter by employing G2, VA4 and NLZ models.  Following Refs.~\cite{horowitz91,niembro01}, we start with the neutrino differential scattering cross-section  
 \begin{eqnarray}
\frac{1}{V}\frac{d^{3} \sigma}{d^{2}{\Omega}'dE'_{{\nu}}} &=& -\frac{G_{F}}{32{\pi}^{2}}\frac{E'_{{\nu}}}{E_{{\nu}}}{\rm Im}(L_{{\mu}{\nu}}{\Pi}^{{\mu}{\nu}}). 
\end{eqnarray}
Here $E_{{\nu}}$ and $ E'_{{\nu}}$ are the initial and final neutrino energies, respectively,  $G_{F}= 1.023{\times} 10^{-5}/M^{2}$ is the weak coupling, and $M$ is the nucleon mass. The neutrino tensor $L_{\mu\nu}$ can be written as 
\begin{eqnarray}
L_{{\mu}{\nu}} &=& 8[2k_{{\mu}}k_{{\nu}}+(k.q)g_{{\mu}{\nu}}-(k_{{\mu}}q_{{\nu}}+q_{{\mu}}k_{{\nu}})\nonumber\\
&{\mp}&{\it i}{\epsilon}_{{\mu}{\nu}{\alpha}{\beta}}k^{{\alpha}}q^{{\beta}}],
\end{eqnarray}
where $k$ is the initial neutrino four-momentum and $q=(q_{0},{\vec{q}})$ is the four-momentum transfer. The polarization tensor ${\Pi}^{{\mu}{\nu}}$, which defines the target particle species, can be written as 
\begin{eqnarray}
{\Pi}^{\it j}_{{\mu}{\nu}}(q) &=& -i \int\frac{d^{4}p}{(2{\pi})^{4}}{\rm Tr}[G^{\it j}(p)J^{\it j}_{{\mu}}G^{\it j}(p+q)J^{\it j}_{{\nu}}],
\end{eqnarray}
where ${\it j}$= $n, p, e^{-}, {\mu}^{-}$. $G(p)$ is the target particle propagator and  $p=(p_0,{\vec{p}})$ is the corresponding initial four-momentum. The currents $J^{\it j}_{{\mu}}$ are $\gamma^{\mu}(C_V^j-C_A^j \gamma_5)$. The explicit forms of $G^j(p)$, $C_V^j$ and $C_A^j$ of every constituent and also their explanations can be found in Ref.~\cite{niembro01}. The NMFP (symbolized by ${\lambda}$) as a function of the initial neutrino energy at a certain density is obtained by integrating the cross section over the time- and vector-component of the neutrino momentum transfer. As a result we obtain~\cite{horowitz91,niembro01} 
\begin{eqnarray}
\frac{1}{{\lambda}(E_{{\nu}})} = \int_{q_{0}}^{2E_{{\nu}}-q_{0}}d|{\vec{q}}|\int_{0}^{2E_{{\nu}}}dq_{0}\frac{|{\vec{q}}|}{E'_{{\nu}}E_{{\nu}}}
2{\pi}\frac{1}{V}\frac{d^3{\sigma}}{d^2{\Omega}'dE'_{{\nu}}}.\nonumber\\
\end{eqnarray}

\begin{figure}[t]
\centering
\begin{tabular}{c@{\qquad}c}
 \mbox{\epsfig{file=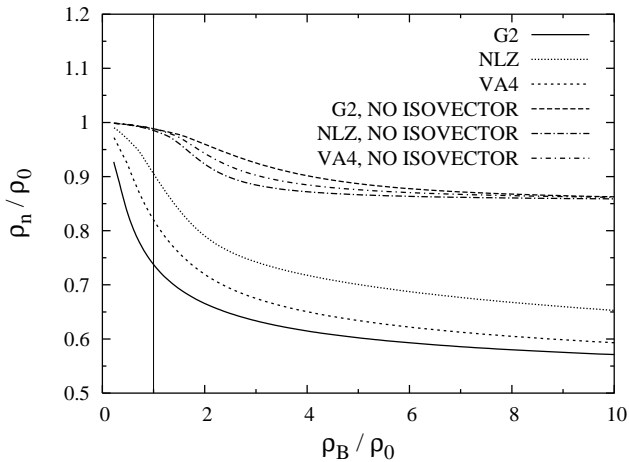,height=6.2cm}}
  \end{tabular}
\caption{Neutron fraction in the neutron star matter with and without isovector terms.}\label{fig2}
\end{figure}

Since in our study we assume that the neutron star matter consists only of neutrons, protons, electrons, and muons, the relative fraction of each constituent should be taken into account in the NMFP calculation. The relative fraction is determined by the chemical potential equilibrium and the charge neutrality of the neutron star at zero temperature. The neutron fractions for all models are shown in Fig.~\ref{fig2}.
 
Qualitatively, all parameter sets have similar trend in fraction of each constituent, i.e., when the neutron fraction is decreasing, other constituent ($p, e^{-}, {\mu}^{-}$) fractions are increasing. Quantitatively,
isovector terms are responsible for the high proton fraction.
 G2 has smaller neutron fraction than VA4 and NLZ. Therefore, even though G2 has an acceptable EOS, it has a too large proton fraction. This fact leads to such a low threshold density for direct URCA process. We note that this fact is ruled out by the analysis of the neutron stars cooling data~\cite{blaschke,yako,tsura}. Thus, this result indicates that significant improvements in the treatment of isovector sector of ERMF-FR are urgently required. Variational calculation of Akmal \etal~\cite{akmal} allows for a direct URCA process only for  ${\rho_B}/{\rho_0}$ $>$ 5. Linear Walecka (linear FR) and Zimanyi-Moszkowski (derivative coupling) Hartree-Fock models of Ref.~\cite{niembro01} yield a higher critical density for the direct URCA process. Isovector contributions of these models do not drastically change the proton fraction.  But on the other hand, all Hartree-Fock models of Ref.~\cite{niembro01} are unable to give a good prediction in finite nuclei, especially in the single particle properties~\cite{quelle,berna}. It may be interesting to see also the consistency of their EOS with experimental data~\cite{daniel}.  
\begin{figure}[t]
\centering
\begin{tabular}{c@{\qquad}c}
 \mbox{\epsfig{file=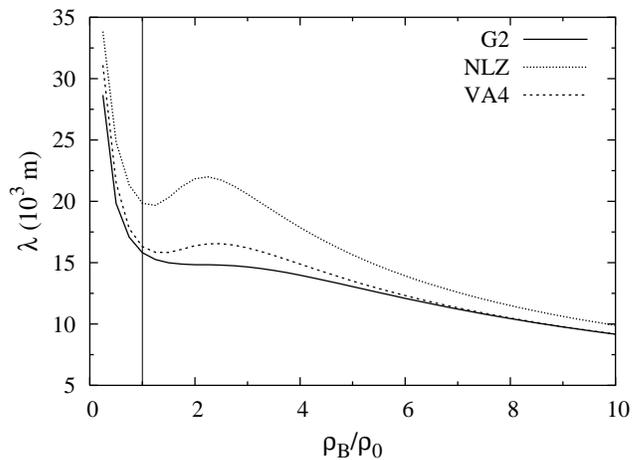,height=6.2cm}}
\end{tabular}
\caption{Neutrino mean free path (NMFP) in the neutron star matter.}\label{fig4}
\end{figure}


\begin{figure}[t]
\centering
\begin{tabular}{c@{\qquad}c}
 \mbox{\epsfig{file=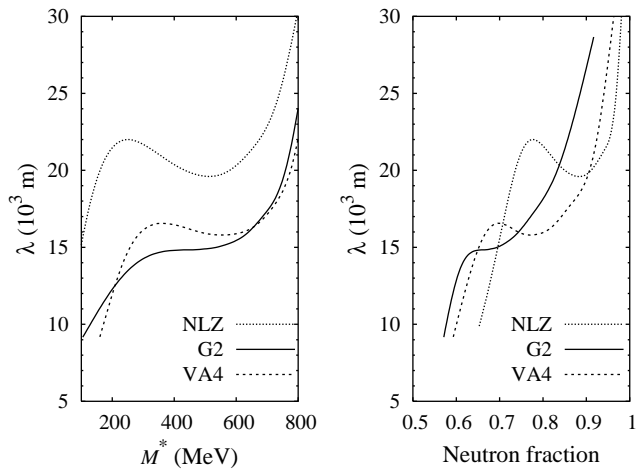,height=6.3cm}}
  \end{tabular}
\caption{Neutrino mean free paths (NMFP) in the neutron star matter as functions of $M^*$ and neutron fraction for all parameter sets.}\label{fig7}
\end{figure}

The NMFP for all models can be seen in Fig.~\ref{fig4}. Here we use a neutrino energy of $E_{\nu}$ = 5 MeV. In general, from medium to high density, $\lambda_{NLZ}$ is larger than $\lambda_{G2}$ and $\lambda_{VA4}$. In high density region we clearly see that  $\lambda_{G2}$  $\approx$ $\lambda_{VA4}$. The NMFP difference among all models appears to be significant around $1\leq {\rho_B}/{\rho_0}\leq 5$ (medium density). For ${\rho_B}/{\rho_0}$ smaller than 1, $\lambda_{VA4}$ $\approx$  $\lambda_{G2}$  $\approx$ $\lambda_{NLZ}$. In other words, in the limit of low density all parameter sets serve similar $\lambda$ prediction as we expected.

In Fig.~\ref{fig7} we show the dependence of $\lambda$ with respect to the $M^*$ and  neutron fraction. Obviously, NLZ has a maximum NMFP at $M^*$ $\approx$ 200 MeV and neutron fraction $\approx$ 0.75. These lead to a bump in $\lambda_{NLZ}$ as shown in Fig.~\ref{fig4}. On the other hand, G2 demonstrates no maximum in $M^*$ 
 and neutron fraction dependences, leading to a smoothly decreasing function of 
$\lambda_{G2}$ displayed in Fig.~\ref{fig4}. For comparison, previous NMFP calculations by using all Hartree-Fock models~\cite{niembro01} showed also no anomaly. In these models, the predicted NMFP falls off faster than that of the Hartree type model as the density increases.

In conclusion, the EOS and NMFP of ERMF models in the high density states have been studied. It is found that the ERMF-FR and ERMF-PC models have different behaviors in high density and  even by using a parameter set that predicts an acceptable EOS, the calculated proton fraction  in neutron star is still too large. Isovector terms are responsible for this. Therefore, improvements in the treatment of the isovector sector of ERMF-FR should be done. Different from the  Hartree-Fock calculation of Ref.~\cite{niembro01}, only the parameter set with an acceptable EOS (G2) has a regular  NMFP.  In order to minimize the anomalous behavior of $\lambda$, a relatively large $M^*$  in RMF models is more favorable. It seems that the relatively large $M^*$ in the ERMF models at high density originates from the presence of the self- and cross-interactions in nonlinear terms. The RMF models with  relatively large $M^*$ retain their regularities partly or fully even for a small neutron fraction.

The works of A.S. and T.M. have been supported in part by the QUE project.

\end{document}